\documentclass[allclo]{FBSart}
\usepackage{graphicx}
\usepackage{amsfonts}
\usepackage{color} 
  \def\nuc#1#2{\relax\ifmmode{}^{#1}{\protect\text{#2}}\else${}^{#1}$#2\fi}
  \def\itnuc#1#2{\setbox\@tempboxa=\hbox{\scriptsize\it #1}
    \def\@tempa{{}^{\box\@tempboxa}\!\protect\text{\it #2}}\relax
    \ifmmode \@tempa \else $\@tempa$\fi}
  
  \newcommand{\beq}{\begin{equation}}
  \newcommand{\eeq}{\end{equation}}
  \newcommand{\bea}{\begin{eqnarray}}
  \newcommand{\eea}{\end{eqnarray}}

\title{The \emph{ab initio} no-core shell model}
\author{C. Forss\'en\instnr{1}\thanks{\textit{E-mail address:} 
christian.forssen@chalmers.se}, 
J. Christensson\instnr{2}, 
P. Navr\'atil\instnr{3},  
S. Quaglioni\instnr{3},
S. Reimann\instnr{2}, 
J. Vary\instnr{4}, 
S. {\AA}berg\instnr{2}
}
\instlist{Fundamental Physics, Chalmers University of Technology, 412 96
  G\"oteborg, Sweden \and  
Mathematical Physics, LTH, Lund University, Box 118, 22 100 Lund,
Sweden \and
Lawrence Livermore National Laboratory, P.O. Box 808, L-414, Livermore,
CA 94551, USA \and
Department of Physics and Astronomy, Iowa State University, %
Ames, IA  50011, USA
}  
\runningauthor{C. Forss\'en et al.}
\runningtitle{The ab initio no-core shell model}
\begin{document}

\maketitle
\begin{abstract}
This contribution reviews a number of applications of the \emph{ab
  initio} no-core shell model (NCSM) within nuclear physics and beyond. 
We will highlight a nuclear-structure study of the $A=12$ isobar using a
chiral NN+3NF interaction.
In the spirit of this workshop we will also mention the new development
of the NCSM formalism to describe open channels and to approach the
problem of nuclear reactions.
Finally, we will illustrate the universality of the many-body problem by
presenting the recent adaptation of the NCSM effective-interaction
approach to study the many-boson problem in an external trapping
potential with short-range interactions.
\end{abstract}

\paragraph{\bf Introduction.}
A truly first-principles approach to the nuclear many-body problem
requires a nuclear Hamiltonian that is based on the underlying theory of
QCD. A candidate for providing the desired connection between QCD and
the low-energy nuclear physics sector is chiral perturbation theory
($\chi$PT), see, e.g., the review by E. Epelbaum~\cite{epe06:57} and
references therein. A very interesting observation from $\chi$PT is that
three-nucleon forces (3NF) appear naturally already at the
next-to-next-to leading order of the expansion. This chiral 3NF was
recently implemented in nuclear many-body calculations as will be
discussed in the next section.

\begin{figure}[htb]
  \centering
  \includegraphics*[width=0.66\columnwidth]
      {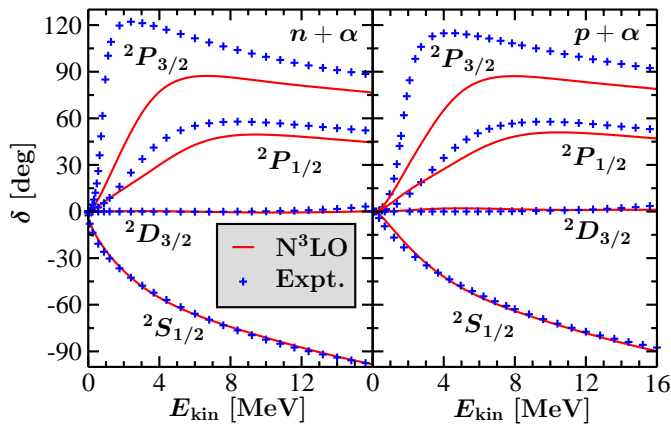}\\
      \caption{Phase shifts for $n$-$\alpha$ (left panels) and
        $p\,$-$\alpha$ (right panels) scattering. Recent NCSM/RGM
        results compared to
        an $R$-matrix analysis of experimental data. From
        Ref.~\cite{qua08:101}.%
        \label{fig:scat}}
\end{figure}

Regardless of its origin, high-precision nuclear Hamiltonians are very
difficult to implement when solving the nuclear many-body problem. 
At this workshop we have heard about a number of methods that are
available to solve the few-body problem ($A = 3 - 4$) to basically
numerical precision.
For more than four particles there are only a handful of methods
available when using modern, realistic interactions. 
Much effort has been spent in studying different unitary transformations
of the interaction to make it tractable for actual many-body
calculations. In particular, the \emph{ab initio} no-core shell model
(NCSM) is usually combined with the cluster-approximated, Lee-Suzuki
transformation to generate effective interactions, see e.g.,
Refs.~\cite{ncsm}.
%
%
%
%
In short, the NCSM is a general approach for studying strongly
interacting, quantum many-body systems. It's a matrix diagonalization
technique to solve the translational invariant $A$-body problem in a
finite harmonic oscillator basis. A particularly nice feature of the
method is the flexibility of the harmonic-oscillator model space that
implies basically no restrictions regarding the choice of
Hamiltonian. Specifically, the NCSM method allows to test the modern
$\chi$PT interactions in many-body calculations.
\paragraph{\bf Recent NCSM Results.%
\label{sec:results}}
The $A=12$ nuclear systems provide a challenge for modern \emph{ab
  initio} methods. The systems can potentially act as new benchmarks as
relevant observables allow for sensitive tests of the nuclear
Hamiltonians and the computed wave functions.
The current level of our experimental understanding of \nuc{12}{C}
includes two bound states and the
triple-alpha threshold at 7.3~MeV. Above this the picture becomes very
complicated due to overlapping broad resonances. A central question
concerns the possible existence of broad $0^+$ and $2^+$ resonances in
this region.  
An important concept that attracts much theoretical interest is the
interplay between triple-alpha and neutron-proton degrees of
freedom.
Studies of ground- and excited states in $A=12$ systems are possible
within the NCSM. These studies are particularly interesting since
the chiral 3NF was recently implemented by P. Navr\'atil et
al.~\cite{nav07:99}. The inclusion of these terms in the NCSM gives the
correct ordering of $T=1$ states with the isobaric analogue of the
\nuc{12}{B} and \nuc{12}{N} ground states being the lowest. It also
provides the correct ordering of the $1^+$ and $4^+$ states although it
over-corrects the spin-orbit strength~\cite{nav07:99}.
Still, regardless of the interaction being used, these results
demonstrate a limitation of the NCSM method. Whereas the spectrum
and properties of shell-model like states are reproduced very nicely,
states that are known to exhibit a high degree of clusterization are
missing from the low-energy spectrum. They typically end up at much
higher excitation energy and are far from converged.
%
\paragraph{\bf Open quantum systems.%
\label{sec:oqs}}
A long-term vision for nuclear theory is to achieve a unified picture of
the nuclear many-body system, including both bound and continuum states
and the transitions between them. Preferably this picture should be
grounded in the fundamental interactions between the constituent
nucleons. In addition, the separation of scales known to occur in
nuclear systems, should be properly described. This requires the
simultaneous modeling of small-scale many-body degrees of freedom and
large-scale few-body correlations.
A possible route towards achieving such a microscopic picture of open
channels and nuclear reactions is explored at Livermore by combining the
NCSM formalism with resonating group methods (RGM)~\cite{qua08:101}.
In the RGM approach the many-body wave function is decomposed into
contributions from various channels that are distinguished by their
different arrangement of the nucleons into clusters.
By defining a set of antisymmetrized cluster basis functions, and
diagonalizing the Hamiltonian in this space, one obtains a non-local,
coupled-channels Schr\"odinger Equation for the relative motion of the
clusters in the different channels.
In Ref.~\cite{qua08:101} this approach was implemented and tested for
certain $A=4-5$ low-energy, single-nucleon scattering problems. In
particular, $n+$\nuc{4}{He} scattering at low energies represents a
convenient training ground for many-body scattering calculations. There
is no $A=5$ bound state, and single-channel scattering is valid up to
rather high energies. There is a sharp, low-energy resonance in the
$3/2^-$ channel, and a broader, high-energy resonance in the $1/2^-$
channel. Scattering in the $s$-wave channel is non-resonant but
obviously depends critically on proper antisymmetrization. Phase shifts
for both $n+$\nuc{4}{He} and $p+$\nuc{4}{He} scattering, calculated in
the NCSM/RGM approach, are presented in Fig.~\ref{fig:scat}.
The method shows very good convergence behavior, but it's clear that
the position and widths of the $p$-wave resonances depend sensitively on
the interaction model.
%
%
\paragraph{\bf Effective Interaction Approach to the Many-Boson Problem.%
\label{sec:boson}}
The emerging field of cold-atom physics has proven to be a very rich
arena of research for few- and many-body physicists. Particle numbers
can be varied, the interaction strength can in many cases be tuned
through Feshbach resonances, and many different properties can be
studied very cleanly in the laboratory.
\begin{figure}[htb]
  \centering
  \includegraphics*[width=0.66\columnwidth]
      {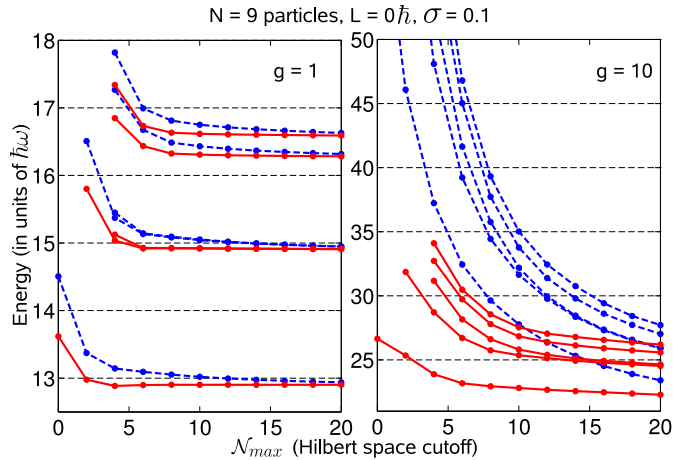}\\
      \caption{Energies for a system of nine bosons and total angular
        momentum $L=0$, for different many-body space cutoffs
        ($\mathcal{N}_\mathrm{max}$). Repulsive Gaussian interactions with range
        $\sigma=0.1$ and two different strengths ($g$) are used
        (oscillator units). The blue-dashed (red-solid) curves
        correspond to standard CI (effective
        interaction approach) calculations. From
        Ref.~\cite{chr08:0802.2811}. %
        \label{fig:N9}}
\end{figure}
Nuclear physics techniques and tools have proven to be very useful to
describe the physics of these systems. With trapping potentials that are
very close to harmonic, the NCSM should be a perfect method. We recently
adapted the NCSM formalism to describe a two-dimensional system of
strongly interacting bosons~\cite{chr08:0802.2811}. A purely repulsive,
short-ranged interaction was modeled with a Gaussian potential. Note
that the different statistics of the bosonic many-body system required a
complete rewrite of the NCSM suite of codes.

The success of the NCSM effective-interaction approach is demonstrated in
Fig.~\ref{fig:N9}. Ground- and excited-state energies are presented for
a system of nine atoms. The NCSM results are compared to the much slower
convergence of the standard configuration interaction (CI) method.
The figure illustrates that stronger correlations within the system are
obtained when increasing the interaction strength (right panel).  In
this case, the computed energies still show a slow decrease with
increasing model space ($\mathcal{N}_\mathrm{max}$). Still, in
comparison, the energies obtained from the standard CI calculations show a
much slower convergence.
These results represent an important first step of our new
approach. Three-dimensional systems and higher particle numbers should
also be within reach for future studies.
\paragraph{\bf Conclusion.%
\label{sec:concl}}
Recent applications of the \emph{ab
  initio} NCSM within nuclear physics and beyond have been reviewed. In
particular, we have demonstrated the study of chiral 3NF Hamiltonians in
the p-shell, the treatment of open channels using the NCSM/RGM approach,
and the effective-interaction approach to the many-boson problem.
%
\begin{acknowledge}
  This research was supported by the Swedish Research Council, the
  Swedish Foundation for Strategic Research, NordForsk, the Knut and
  Alice Wallenberg Foundation, and the U.S. Department of Energy. Partly
  prepared by LLNL under Contract DE-AC52-07NA27344. 
  One of us (C.F.)  acknowledges financial support from Stiftelsen Lars
  Hiertas Minne and Stiftelsen L\"angmanska Kulturfonden.
\end{acknowledge}



\end{document}